\documentclass[aps,prl,preprint,superscriptaddress,showpacs]{revtex4}

\usepackage{epsfig}
\usepackage{graphicx}
\usepackage{latexsym}

\usepackage{color}

\begin{document}

\title{Spiral Growth and Step Edge Barriers}

\author{Alex Redinger}
\affiliation
{II. Physikalisches Institut, Universit{\"a}t zu K{\"o}ln, 50937 K{\"o}ln, Z{\"u}lpicher Str. 77,
Germany}
\affiliation
{I. Physikalisches Institut, RWTH Aachen, 52056 Aachen,Germany}
\author{Oliver Ricken}
\affiliation
{II. Physikalisches Institut, Universit{\"a}t zu K{\"o}ln, 50937 K{\"o}ln, Z{\"u}lpicher Str. 77,
Germany}
\author{Philipp Kuhn}
\affiliation
{Institut f{\"u}r Theoretische Physik, Universit{\"a}t zu K{\"o}ln, 50937 K{\"o}ln, Z{\"u}lpicher Str. 77,
Germany}
\author{Andreas R{\"a}tz}
\affiliation{Crystal Growth Group, Research Center Caesar, Ludwig-Erhard-Allee 2, 53175 Bonn, Germany}
\author{Axel Voigt}
\affiliation{Crystal Growth Group, Research Center Caesar, Ludwig-Erhard-Allee 2, 53175 Bonn, Germany}
\affiliation{Institut f\"ur Wissenschaftliches Rechnen, Technische Universit\"at Dresden, Zellescher Weg
12 --14, 01062 Dresden, Germany}
\author{Joachim Krug}
\affiliation
{Institut f{\"u}r Theoretische Physik, Universit{\"a}t zu K{\"o}ln, 50937 K{\"o}ln, Z{\"u}lpicher Str. 77,
Germany}
\author{Thomas Michely}
\affiliation
{II. Physikalisches Institut, Universit{\"a}t zu K{\"o}ln, 50937 K{\"o}ln, Z{\"u}lpicher Str. 77,
Germany}

\date{\today}

\begin{abstract}

The growth of spiral mounds containing a screw dislocation is compared to the growth 
of wedding cakes by two-dimensional nucleation. Using phase field simulations and homoepitaxial growth experiments on the 
Pt(111) surface we show that both structures attain the same characteristic large scale shape when a significant step edge
barrier suppresses interlayer transport. The higher vertical growth rate observed for the spiral
mounds on Pt(111) reflects the different incorporation mechanisms for atoms in the top region and can be formally represented
by an enhanced apparent step edge barrier.   

\end{abstract}


\maketitle
When a screw dislocation intersects a crystal surface, a step forms which emerges from the intersection
point and cannot be removed by growth or evaporation. Under deposition the step turns around the intersection
point, resulting in a characteristic spiral mound. This mechanism was conjectured by Frank in 1949
to account for the enormous discrepancy between experimentally observed crystal growth rates and the 
predictions of two-dimensional nucleation theory \cite{Burton49,BUR51}. Since then, spiral mounds have 
been recognized as a ubiquitous feature
of many growth systems ranging from high temperature superconductors \cite{Gerber91}
to biominerals \cite{Teng98} and organic thin films \cite{Ruiz04}. 

Theoretically, the prediction of the shape of a growth spiral involves 
a moving boundary value problem for the adatom concentration field in a complex geometry
\cite{Markov04},
which has been treated by approximate analytical calculations
\cite{Surek73,vanderEerden81} as well as by numerical phase field techniques \cite{Karma98}. 
These approaches result in a simple conical mound shape with a constant step spacing,
which is proportional to the step radius of curvature near the spiral core
and generally decreases as a power law with increasing deposition flux. 

In the growth of organic thin films frequently spiral mounds are observed whose shapes are inconsistent with 
this classical scenario. They possess small flat plateaus at the top, a slope (or step spacing) that varies
non-monotonically with the distance from the spiral core, and they 
are typically separated by deep grooves \cite{Ruiz04,Beigmohamadi07}. These shapes are  
strikingly reminiscent of the mound structures, often referred to as 
\textit{wedding cakes}, that form during growth 
on dislocation-free metal surfaces \cite{Krug97,Politi97,Kalff99,Michely04}
when the mass transport between different layers is suppressed
by a strong step-edge barrier. 
It is then plausible to hypothesize that the unconventional spiral mounds
observed on organic thin films may be related to a step-edge barrier,
which has been conjectured to exist on the basis
of various indirect pieces of evidence \cite{Ruiz04,Krause04,Zorba06,Mayer06}. 

In the present work we therefore investigate experimentally and theoretically the hitherto disregarded situation of 
spiral growth in the presence of a step edge barrier. Our starting point is the observation  
that an atom finding itself on the hillside of a spiral mound,
far away from the core, diffuses in an environment that is indistinguishable from that on the hillside of a wedding cake
consisting of concentric circular islands \cite{Surek73}. 
The overall shape of the spiral mound must therefore reflect the strength of the 
step edge barrier in the same way as for a wedding cake. 

\begin{figure}
\begin{center}
\includegraphics[width=1.0 \linewidth]{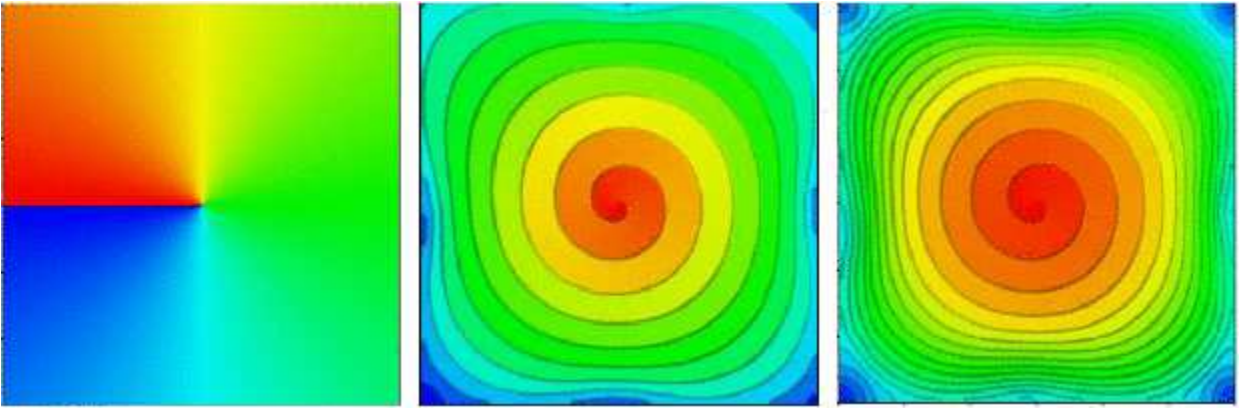}
\caption{(Color online) Growth spirals obtained from phase field simulations with 
$D = 10$, $F = 0.2$, $\gamma = 1$,
$\rho^\ast = 0.1$, $k_+ = 10$, $k_- = 1$. Panels show the surface before growth (left) and after deposition 
of 140 ML (middle) and 300 ML (right), respectively.} 
\label{Fig_Phasefield}
\end{center}
\end{figure}

This is illustrated in 
Fig. \ref{Fig_Phasefield} for a growth spiral
generated numerically using a phase field model \cite{Otto04,Raetz07}.
The model is a diffuse interface approximation of the Burton-Cabrera-Frank (BCF) theory \cite{BUR51},
which takes the form of a moving boundary value problem for the adatom concentration on the surface.
Key parameters of the model are the surface diffusion coefficient $D$ and the deposition flux $F$, 
as well as the kinetic coefficients $k_\pm$ governing the attachment of atoms to a step from the
lower ($k_+$) and upper ($k_-$) terrace, respectively. Step thermodynamics enters through 
the step stiffness $\gamma$ and the equilibrium adatom concentration $\rho^\ast$.  
In the phase field approximation the crystal surface is represented by
a continuous field $\phi(\vec r,t)$ which couples to the adatom
concentration and is subjected to a 
multiwell potential that pins $\phi$ to the preferred
integer values representing the discrete surface heights. 
The step edge barrier is incorporated through a mobility function
that is minimal at the steps and
asymmetric with respect to half-integer values of $\phi$.
In the numerical implementation the model is non-dimensionalized
by scaling lengths and times with suitable microscopic quantities.   

A screw dislocation is
introduced by the shift $\phi \to \phi - \theta(\vec r)/2 \pi$, where
$\theta(\vec r)$ is the planar rotation angle around the dislocation core \cite{Karma98,Raetz07}.
In the absence of a step edge barrier the simulation of the model converges to a conical mound with a
spatially constant, time-independent slope. By contrast, the spiral mound displayed in Fig.~\ref{Fig_Phasefield}
has a flat top and curved hillsides which steepen indefinitely with increasing deposition flux, up to the point where
further steepening is limited by the intrinsic step width of the phase field. This is precisely the behavior expected
from the analogy with wedding cakes \cite{Krug97,Michely04}. 

Despite the similarity in the large scale shape, it is clear that
the growth mechanism near the top of the
structure is completely different for spiral mounds and for wedding cakes. 
Even the answer to the simple question as to which of the two 
grows faster under a given set of conditions is not obvious. The growth of wedding cakes is governed by the rate of 
two-dimensional nucleation
of islands on the top terrace, which is enhanced by the confinement of atoms due to the step edge barrier \cite{Krug00}. 
In contrast, atoms landing near the top of the spiral mound 
can avoid the step edge barrier by moving around the core, which should \textit{reduce} the local adatom concentration
and hence the growth rate. On the other hand, the indelible presence of the step emanating from the screw
dislocation obviates the need for two-dimensional nucleation, which should \textit{increase} the rate of vertical growth.
When the adatom concentration in our simulations     
was allowed to become sufficiently large so that islands nucleated
deterministically, spiral mounds were generally seen to grow higher than regular wedding cakes \cite{Raetz07},
indicating that the second effect dominates. 

\begin{figure}
\begin{center}
\includegraphics[width=0.7 \linewidth]{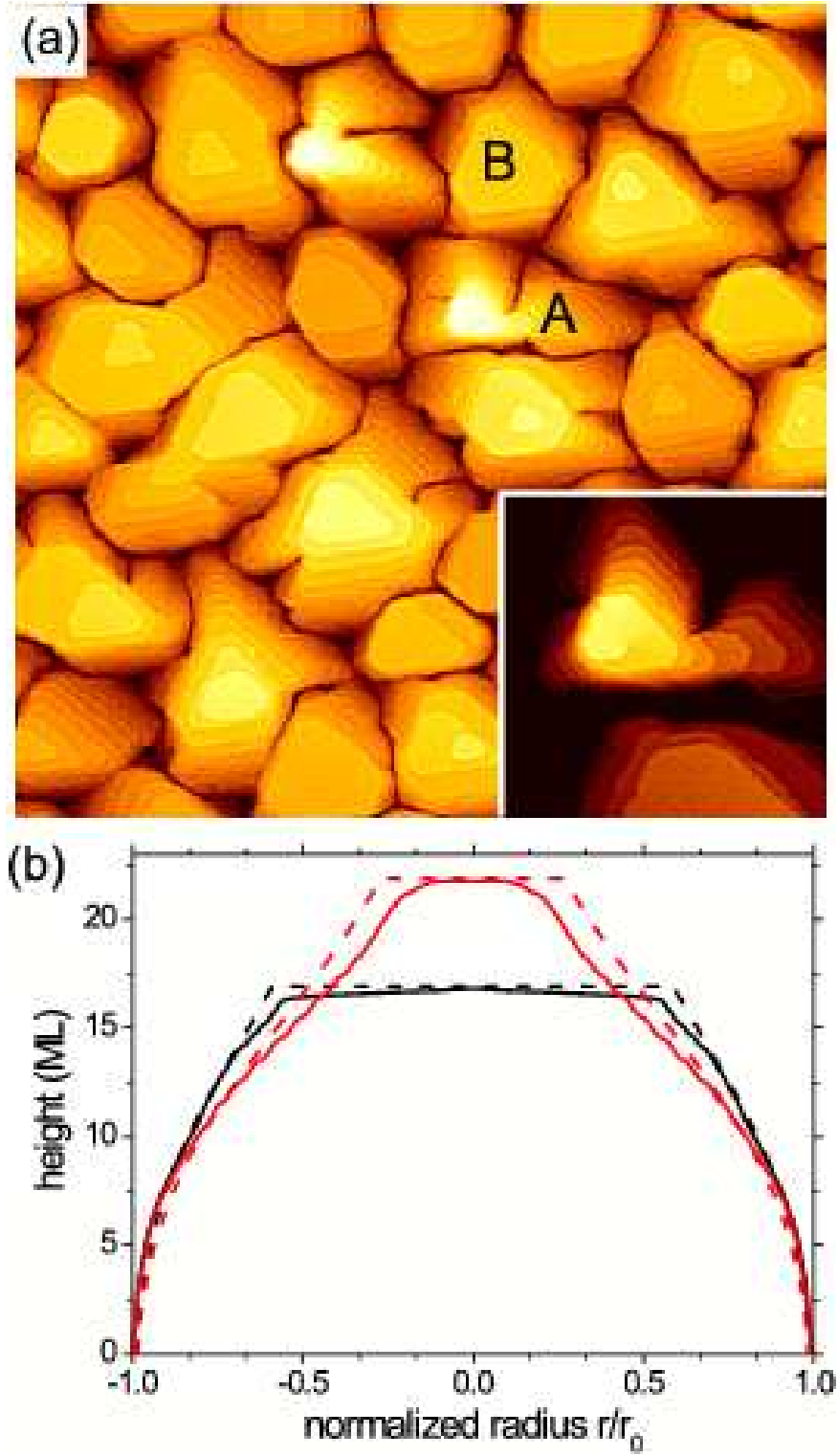}
\caption{(Color online) (a) Morphology of Pt(111) after deposition of 39\,ML at 400\,K on a substrate area 
with screw dislocation intersection points. Image size 1750\AA\,$\times$\,1750\AA. Inset: Enlarged view of the 
top area of the spiral mound labeled A. (b) Radially averaged height distributions of the spiral mound 
labeled A in (a) [full red/gray line] and the wedding cake labeled B in (a) [full black line]. 
The fits to the shape profiles according to the 
analytic model are plotted as dashed lines.}
\label{Fig_Topographs}
\end{center}
\end{figure}

To experimentally compare the two growth mechanisms under identical conditions, we have devised a  setup in which
they are \textit{simultaneously} realized on the Pt(111) surface.
The experiments were performed in an ultra high vacuum variable temperature scanning tunneling microscopy (STM) apparatus with a base pressure in the $10^{-11}$\,mbar range \cite{Bott951}. Cleaning of the 
Pt(111) sample
was accomplished by cycles of flash annealing to 1273 K and sputtering by a mass separated 5\,keV Ar$^+$ ion beam. In order to fabricate a surface with screw dislocation lines intersecting the surface in a suitable areal density, prior to growth the procedure outlined in \cite{Michely91} was applied. Helium was implanted at room temperature into the sample by a 4.5 keV He$^+$ ion beam incident at an angle of $76^{\circ}$ with respect to the surface normal and with a dose of $ 3.8 \times 10^{20}$ ions/m$^2$. Subsequently the surface was annealed for 120\,s to $T=800$\,K. The growth of the precipitating He bubbles gives rise to dislocation loop punching. Dislocation reactions and glide towards the surface results partially in loop annealing (adatom island formation) but partially also in separated pairs of screw dislocation intersection points with the surface (step dipoles). Onto this surface Pt was evaporated with a typical rate of $ 1 \times 10^{-2}$\,ML/s, where 1 monolayer (ML) denotes the surface atomic density of Pt(111). The deposited amounts $\Theta$ range from 1 to 100\,ML and temperatures $T$ from 250 to 500\,K. After deposition the morphology was quenched to avoid changes due to surface diffusion. 
 
Figure \ref{Fig_Topographs} (a) represents the morphology resulting after deposition of a total coverage $\Theta $
of 39 ML at 400\,K. 
As in previous experiments on dislocation-free surfaces \cite{Kalff99,Michely04}, a large number of mounds are formed that exhibit 
flat tops and are separated by deep grooves. Two of the mounds visible in Fig. \ref{Fig_Topographs} (a) are significantly higher than all others; their heights 
exceed the surrounding mounds by 4-6 ML. Closer inspection reveals that both of them carry a screw dislocation intersection point 
in their center, as exemplified by the inset. Figure \ref{Fig_Topographs} (b) displays the mound profiles obtained from the height 
distributions of the spiral mound labeled A and the wedding cake labeled B in Fig. \ref{Fig_Topographs} (a). The profiles 
are obtained by redistributing the layer coverages into a stack of circular islands.
Both mounds clearly show flat top plateaus. The spiral mound profile displays in addition a change from positive curvature close to the top to negative curvature close to the edge. 
Such an inflection point in the height profile is characteristic for mounds grown in the presence of large step edge barriers \cite{Michely04}. 

The dashed lines in Fig. \ref{Fig_Topographs} (b) represent fits to the function
$
\theta(h) = 1 - C\{1 + \mathrm{erf}[(h-\Theta)/\sqrt{\Theta}]\},
$
where $\theta(h)$ is the fractional coverage at height $h$ and 
erf denotes the error function. The width of the mound at height $h$
is proportional to $\sqrt{\theta(h)}$. This equation follows
from a simple model for the growth of wedding cakes \cite{footnote2}, in 
which mass transport between all layers except the top one is completely 
suppressed and a new top terrace nucleates when the current one
has reached a critical coverage $\theta_c$ \cite{Michely04}. The only fit parameter $C$ 
is related to $\theta_c$ and is most simply determined from the 
total height of the mound. 
While the agreement with the experimental data is not perfect,
the fit is satisfactory in view of the crudeness of the model and the 
large number of factors that have been neglected in its application.
For example, repulsive step-step interactions come into play when the steps on the 
hillside approach each other and tend to reduce the mound slope, as observed in
Fig.\ref{Fig_Topographs} (b).   

Via the theory of second layer nucleation \cite{Michely04,Krug00},
the critical coverage $\theta_c$ 
can be linked to the additional step edge barrier 
$\Delta E_{\rm S}$, which is defined as the difference between the activation energy
for a descending hop into the step position compared to a hop within the layer \cite{footnote1}.
A large value of $\Delta E_{\rm S}$ enhances the confinement of adatoms on the top terrace, 
which increases the probability for nucleation, decreases $\theta_c$, and increases
the height of the mound.  
Using the known data for adatom diffusion on Pt(111) \cite{Michely04}, the fits shown in Fig.\ref{Fig_Topographs} (b) translate into
the estimates  $\Delta E_{\rm S}=$ 0.22 eV for the wedding cake (mound B) and $\Delta E_{\rm S}=$ 0.34 eV for the spiral mound (mound A). 
Although spiral mounds obviously do not grow by two-dimensional nucleation, we can formally represent the experimentally observed faster
growth of the spiral mounds by an increase of the \textit{apparent} step edge barrier by 0.12\,eV.

\begin{figure}
\begin{center}
\includegraphics[width=0.8 \linewidth]{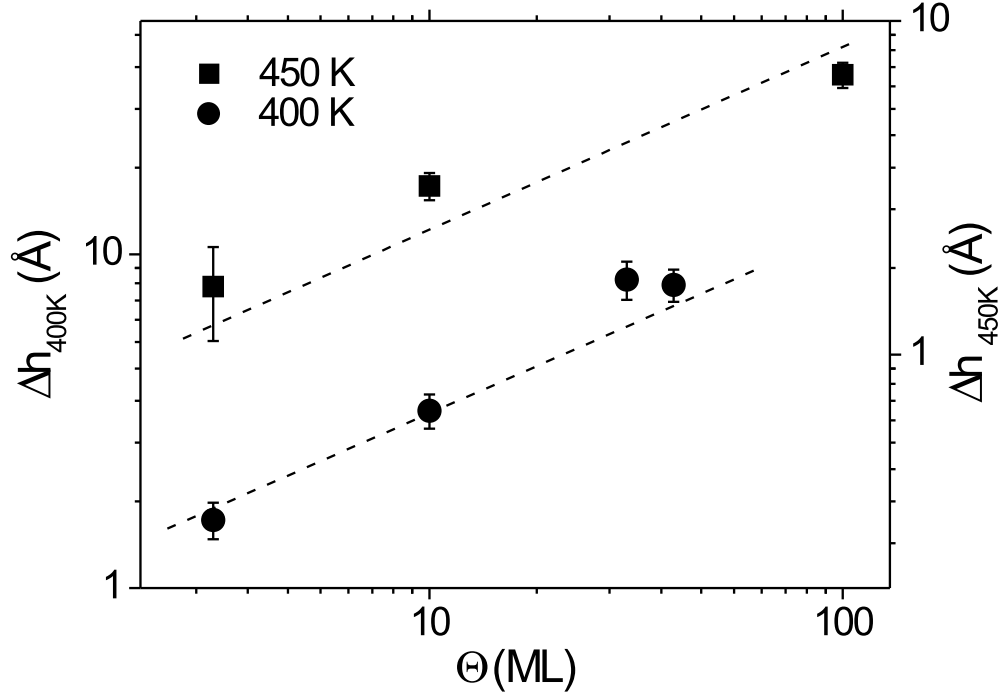}
\caption{Average height difference $\Delta h$ between spiral mounds and wedding cakes for 400\,K (full circles) and 450\,K (full squares) as a function of $\Theta$.
The dashed lines correspond to $\Delta h \propto \sqrt{\Theta}$ as predicted by the mound growth model.}
\label{Fig_Heights}
\end{center}
\end{figure}

Another prediction of the model is that the height difference between two mounds characterized by different values of 
$C$ or $\theta_c$ grows proportional to $\sqrt{\Theta}$. This prediction should then also
apply to the average height difference $\Delta h$ between spiral mounds and wedding cakes.
Figure \ref{Fig_Heights} shows experimental data for $\Delta h$ obtained by measuring the extra height of a spiral mound with respect to the heights of the wedding cakes being its direct neighbors. 
This quantity was then averaged for each temperature $T$ and coverage $\Theta$ over an ensemble of spiral mounds. 
Indeed Fig. \ref{Fig_Heights} shows that $\Delta h$ grows roughly proportional to $\sqrt{\Theta}$ for the two temperatures analyzed.

\begin{figure}
\begin{center}
\includegraphics[width=0.9 \linewidth]{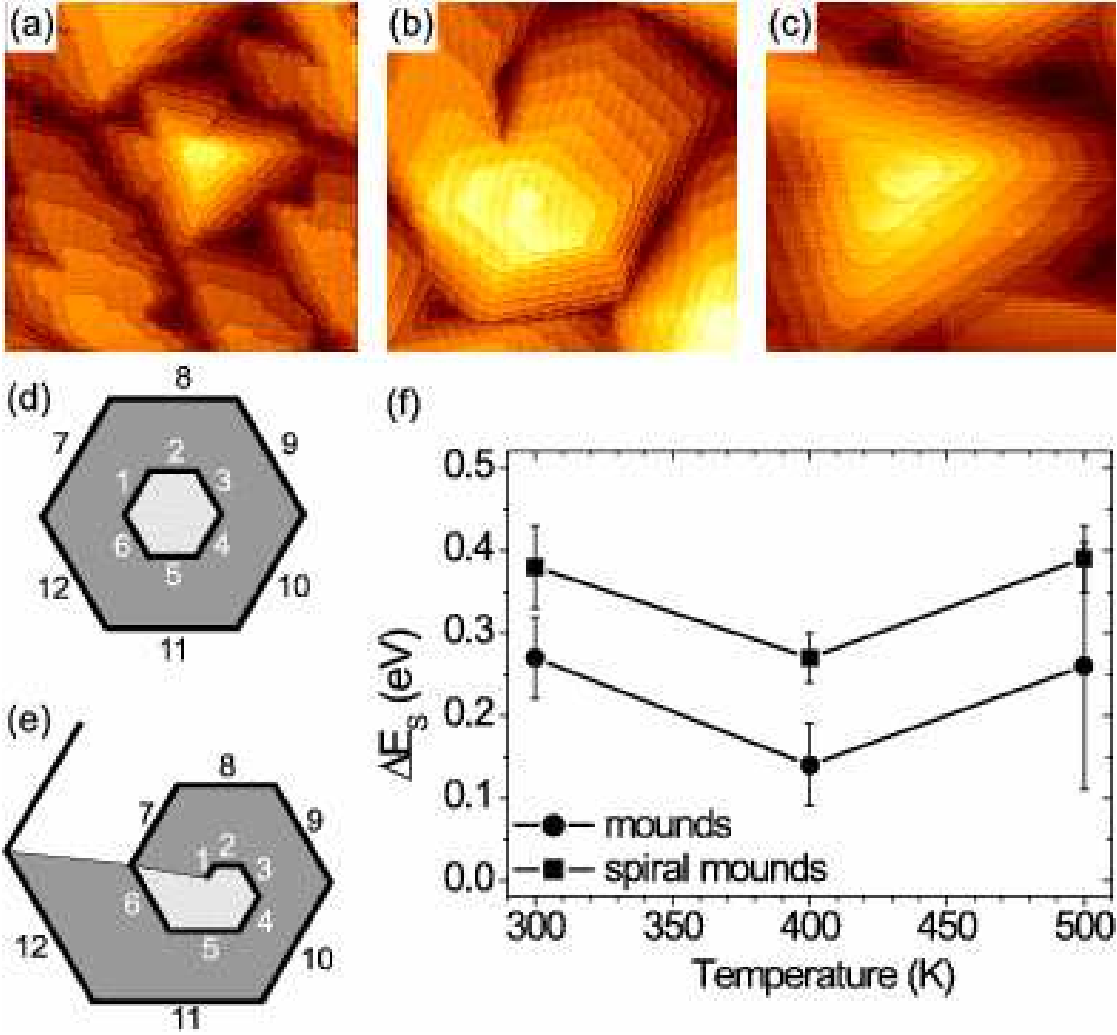}
\caption{(Color online) 
Examples of spiral mounds after deposition of 10\,ML at (a) 300\,K, (b) 400\,K, (c) 500\,K. Image size is always 400\AA\,$\times$\,400\AA. Schematic top view sketch of the top area of (d) 
a wedding cake and (e) a spiral mound. The numbers label the step segments. (f) Results for $\Delta E_{\rm S}$ for wedding cakes (full dots) for the three growth temperatures investigated together with the 
apparent $\Delta E_{\rm S}$ for spiral mounds (full squares) grown simultaneously.}
\label{Fig_TopTerraces}
\end{center}
\end{figure}

As a further test of the concept of an increase in the apparent step edge barrier for spiral mounds, we have 
determined the sizes of the flat top region for mounds grown at various temperatures with a total coverage of  
10\,ML [see Figs.~\ref{Fig_TopTerraces}(a)-(c)]. To characterize the extension of the top area for spiral mounds, the sum of the lengths of step segments 7 to 12 of the spiral 
is used as sketched in Figs.~\ref{Fig_TopTerraces}(d)-(e). 
This corresponds to the perimeter of the \textit{base terrace}, the terrace below the top terrace of a wedding cake, which is known to show less scatter than 
the top terrace size itself \cite{Krug00}. Using formulae of second layer nucleation theory \cite{Michely04}, the base terrace perimeter translates into estimates for $\Delta E_{\rm S}$ 
for the two kinds of mounds, which are displayed in Fig.~\ref{Fig_TopTerraces}(f). We find a consistent increase of the apparent step edge barrier of spiral mounds by 0.11 - 0.13\,eV, in good agreement
with the result obtained from the mound shapes in Fig.~\ref{Fig_Topographs}.

The results presented so far raise an obvious question: Is the faster growth of spiral mounds observed on Pt(111) a generic feature, or is it specific to this
growth system? To provide at least a qualitative answer, 
note first that only the fate of the adatoms arriving in the top region [the areas shaded light gray in Figs.\ref{Fig_TopTerraces}(d) and (e)] 
determines the shapes and heights of mounds, given that $\Delta E_{\rm S}$ is significant, because then adatoms arriving on the slopes of a mound are incorporated with a probability near unity 
into the ascending step bounding their terrace. In contrast, adatoms arriving near the top of a mound are able to reach a lower layer, either because the top terrace
on which they land is small (in the case of a wedding cake) or because they can move around the spiral core. Thus the area of the top region is directly proportional to the 
effective downhill flux of adatoms. The smaller the top region, the higher the mound and the closer the mound shape is to the statistical growth limit, in which
the layer coverages have a Poisson distribution \cite{Krug97,Michely04}. 

Understanding the difference between the two kinds of mounds therefore amounts to understanding
the physics that determines the size of the top area. For wedding cakes the key process is two-dimensional nucleation \cite{Michely04,Krug00}. For the spiral mounds on Pt(111),
the appropriate physical picture is that of a polygonized spiral growing by the motion of straight segments that follow the crystallographic directions of the surface
\cite{Markov04}. The scale of the top region is then set by the length $l_c$ that the stationary segment 1 in Fig.~\ref{Fig_TopTerraces}(e) has to reach before it can
grow by incorporating adatoms, thus taking over the role of segment 2. Hence $l_c$ is the critical segment length for \textit{one-dimensional} nucleation, and it is determined
by the atomistic processes of step edge diffusion and corner rounding \cite{Michely04}. Applying the kinematic approach of \cite{Markov04} to the fcc(111) surface,
one finds that $l_c$ equals the length of segment 2, which does not change in time. In the present case this yields the estimate $l_c \approx 23 \pm$6\AA,
which is much smaller than the average top terrace size of a wedding cake. 
This discussion makes it clear
that for a system with a different hierarchy of diffusion processes the effect of a spiral step might be qualitatively different. 
If the activation energy for step edge diffusion is small and the step edge barrier is large, 
$l_c$ could be large compared to the size of a top terrace at second layer nucleation, and  
wedding cakes could be higher than spiral mounds. In fact, this has to be the case in the limit
of an infinite step edge barrier, $\Delta E_\mathrm{S} \to \infty$, where the top terrace of a wedding cake
shrinks to a point \cite{Krug97} while the spiral mounds maintain a finite top plateau of size
$\sim l_c$. 


We acknowledge useful discussions with C. Busse, M. Beigmohamadi, A. Farahzadi, P. Niyamakom, 
E.D. Williams, M. Wuttig and W. Kalb.
A. Redinger acknowledges financial support through a Ph. D. Fellowship of RWTH Aachen. 


\begin{thebibliography}{24}
\expandafter\ifx\csname natexlab\endcsname\relax\def\natexlab#1{#1}\fi
\expandafter\ifx\csname bibnamefont\endcsname\relax
  \def\bibnamefont#1{#1}\fi
\expandafter\ifx\csname bibfnamefont\endcsname\relax
  \def\bibfnamefont#1{#1}\fi
\expandafter\ifx\csname citenamefont\endcsname\relax
  \def\citenamefont#1{#1}\fi
\expandafter\ifx\csname url\endcsname\relax
  \def\url#1{\texttt{#1}}\fi
\expandafter\ifx\csname urlprefix\endcsname\relax\def\urlprefix{URL }\fi
\providecommand{\bibinfo}[2]{#2}
\providecommand{\eprint}[2][]{\url{#2}}

\bibitem[{Bur()}]{Burton49}
\bibinfo{note}{W. Burton, N. Carbrera, F. C. Frank, Nature \textbf{163}, 398
  (1949).}

\bibitem[{BUR()}]{BUR51}
\bibinfo{note}{W. K. Burton, N. Carbrera, and F. C. Frank, Philos. Trans. R.
  London Soc. A {\bf 243} (1951) 299.}

\bibitem[{Ger()}]{Gerber91}
\bibinfo{note}{C. Gerber et al., Nature {\bf 350} (1991) 279}.

\bibitem[{Ten()}]{Teng98}
\bibinfo{note}{H.H. Teng et al., Science {\bf 282} (1998) 724}.

\bibitem[{Rui()}]{Ruiz04}
\bibinfo{note}{R. Ruiz et al., Chem. Mater. {\bf 14}, 4497 (2004).}

\bibitem[{Mar()}]{Markov04}
\bibinfo{note}{I.V. Markov \emph{Crystal Growth for Beginners, 2nd edition}
  (World Scientific, Singapore 2004), pp. 218}.

\bibitem[{Sur()}]{Surek73}
\bibinfo{note}{T. Surek, J.P. Hirth, and G.M. Pound, J. Cryst. Growth {\bf 18}
  (1973) 20}.

\bibitem[{van()}]{vanderEerden81}
\bibinfo{note}{J.P. van der Eerden, J. Cryst. Growth {\bf 53} (1981) 305.}

\bibitem[{Kar()}]{Karma98}
\bibinfo{note}{A. Karma and M. Plapp, Phys. Rev. Lett. {\bf 81}, 4444 (1998).}

\bibitem[{Bei()}]{Beigmohamadi07}
\bibinfo{note}{M. Beigmohamadi, P. Niyamakom, A. Farahzadi, S. Kremers, T.
  Michely, M. Wuttig, in preparation.}

\bibitem[{Kru({\natexlab{a}})}]{Krug97}
\bibinfo{note}{J. Krug, J. Stat. Phys. {\bf 87}, 505 (1997).}

\bibitem[{Pol()}]{Politi97}
\bibinfo{note}{P. Politi, J. Phys. I {\bf 7}, 797 (1997).}

\bibitem[{Kal()}]{Kalff99}
\bibinfo{note}{M. Kalff, P. \u{S}milauer, G. Comsa, and T. Michely, Surf. Sci.
  {\bf 426}, L447 (1999).}

\bibitem[{Mic({\natexlab{a}})}]{Michely04}
\bibinfo{note}{T. Michely and J. Krug \emph{Islands, Mounds and Atoms, Springer
  Series in Surf. Sci. Vol. 42}, (Springer, Berlin 2004) p. 158 ff.}

\bibitem[{Kra()}]{Krause04}
\bibinfo{note}{B. Krause et al., Europhys. Lett. {\bf 65} (2004) 372.}

\bibitem[{Zor()}]{Zorba06}
\bibinfo{note}{S. Zorba, Y. Shapir, and Y. Gao, Phys. Rev. B {\bf 74} (2006)
  245410.}

\bibitem[{May()}]{Mayer06}
\bibinfo{note}{A.C. Mayer et al., Phys. Rev. B {\bf 73} (2006) 205307.}

\bibitem[{Ott()}]{Otto04}
\bibinfo{note}{F. Otto et al., Nonlinearity {\bf 17} (2004) 477.}

\bibitem[{Rae()}]{Raetz07}
\bibinfo{note}{A. R\"atz, \textit{Modelling and numerical treatment of
  diffuse-interface models with applications in epitaxial growth}, PhD
  dissertation (University of Bonn, 2007).}

\bibitem[{Kru({\natexlab{b}})}]{Krug00}
\bibinfo{note}{J. Krug, P. Politi, and T. Michely, Phys. Rev. B {\bf 61} (2000)
  14037.}

\bibitem[{Bot()}]{Bott951}
\bibinfo{note}{M. Bott, T. Michely, and G. Comsa, Rev. Sci. Instrumen. {\bf
  66}, 4135 (1995).}

\bibitem[{Mic({\natexlab{b}})}]{Michely91}
\bibinfo{note}{T. Michely and G. Comsa, J. Vac. Sci. Technol. B {\bf 9}, 862
  (1991).}

\bibitem[{foo({\natexlab{a}})}]{footnote2}
\bibinfo{note}{The equation applies in the limit of large $\Theta$, where the
  fact that the height distribution of a wedding cake is discrete, in contrast
  to the continuous distribution of a spiral mound, is irrelevant.}

\bibitem[{foo({\natexlab{b}})}]{footnote1}
\bibinfo{note}{We use $\Delta E_{\rm S}$ as effective barrier averaging over
  the atomistic details of the step configuration, which in fact give rise to a
  variation of the barrier with step atomic structure and also neglect details
  of the potential energy landscape at the step edge (compare
  \cite{Michely04}).}

\end{thebibliography}

\end{document}